\documentclass[conference]{IEEEtran}
\IEEEoverridecommandlockouts
\usepackage{cite}
\usepackage{amsmath,amssymb,amsfonts}
\usepackage{algorithmic}
\usepackage{graphicx}
\usepackage{textcomp}
\usepackage{xcolor}

\def\BibTeX{{\rm B\kern-.05em{\sc i\kern-.025em b}\kern-.08em
    T\kern-.1667em\lower.7ex\hbox{E}\kern-.125emX}}
    
\usepackage[hyphens]{url}

\usepackage{enumitem}
\usepackage{tikz}
\usepackage{booktabs}
\usepackage{array}
\usepackage{tabularx}

\usepackage[bookmarks=true,breaklinks=true,colorlinks,linkcolor=black,citecolor=blue,urlcolor=black]{hyperref}

\definecolor{color1}{RGB}{181, 2, 2}
\providecommand{\keywords}[1]
{\small \textbf{\textit{Keywords---}} #1}
\title{Zeno: A Scalable Capability-Based Secure Architecture} 

\begin{document}

\author{
\IEEEauthorblockN{Alan Ehret, Jacob Abraham, Mihailo Isakov, Michel A. Kinsy}
\IEEEauthorblockA
    {
        \textit{Secure, Trusted, and Assured Microelectronics (STAM) Center}\\
        \textit{Ira A. Fulton Schools of Engineering, Arizona State University}\\
        Emails: \{aehret1, jrabraha, misakov1, mkinsy\}@asu.edu
    }
}

\maketitle
\pagestyle{plain}

%%%%%% -- PAPER CONTENT STARTS-- %%%%%%%%

\begin{abstract}

Despite the numerous efforts of security researchers,  memory vulnerabilities remain a top issue for modern computing systems.
Capability-based solutions aim to solve whole classes of memory vulnerabilities at the hardware level by encoding access permissions with each memory reference.
While some capability systems have seen commercial adoption, little work has been done to apply a capability model to datacenter-scale systems.
Cloud and high-performance computing often require programs to share memory across many compute nodes. 
This presents a challenge for existing capability models, as capabilities must be enforceable across multiple nodes.
Each node must agree on what access permissions a capability has and overheads of remote memory access must remain manageable.

To address these challenges, we introduce Zeno, a new capability-based architecture. Zeno supports a Namespace-based capability model to support globally shareable capabilities in a large-scale, multi-node system.
In this work, we describe the Zeno architecture, define Zeno's security properties, evaluate the scalability of Zeno as a large-scale capability architecture, and measure the hardware overhead with an FPGA implementation.

\end{abstract}

\section{Introduction}
\label{sec:introduction}

After decades of research, memory related vulnerabilities remain a leading issue for modern systems and software. 
Out-of-bounds-write remains number one on MITRE's Top 25 list of CWEs while use-after-free is number seven \cite{mitre_CWE_top25}.
So far, little research has focused on memory security in high-performance and data-center scale systems.

Simultaneously, efficient data sharing in multi-node systems remains a challenge.
Current solutions use complex software stacks to achieve maximum performance, but these software stacks impart additional latency to each remote data access.
Benchmarks of OpenSHMEM Get operations shows that 50\% of the access latency for a 4-Byte transfer is due to software overhead; for 4kB and 16kB transfers, software overhead remains 25\% and 20\% of the total latency respectively \cite{wang2021xbgas}. 
Memory security solutions that further increase the overhead for node-to-node data sharing will be of limited utility for systems focused on high-performance and scale.

We believe that an opportunity exists to address the both memory security and data sharing challenges faced by modern high-performance and data-center scale systems with a single architecture.
To that end, we introduce \textbf{Zeno} - \textbf{a new capability architecture} with support for global address spaces shareable across nodes. A capability is an unforgeable, protected pointer that includes not only the address, but also an access right specification that is enforced in hardware.  

Capabilities in the Zeno architecture are implemented as Namespaces.
In Zeno, Namespaces are an address space abstraction that is globally shareable in a multi-node system.
All memory is accessed through the Namespace address space abstraction using a base address and Namespace ID.
Hardware uses the Namespace ID to look up metadata associated with the Namespace capability and validate each access.
Address bounds for each Namespace create an accessible range within the address space abstraction.
Access to a Namespace is revokeable by marking the Namespace metadata as invalid.
Namespaces are hierarchical, with parent and children Namespace capabilities providing access to different ranges of the same address space abstraction.
The address space abstraction created by a Namespace capability is decoupled from the physical memory of a multi-node system.
Namespace data may transparently span multiple nodes and remain accessible with the same Namespace capability from any node in the system.
Namespaces in the Zeno architecture make it possible for any data to be globally shareable.
Architecture and microarchitecture support for Namespaces reduce the cost of accessing remote memory, supporting larger pools of shared data and reduced access latency.
The Namespace capability model ensures that, while data is globally sharable, only software with a Namespace capability may access the Namespace data.

The Zeno architecture's support for efficient and secure globally shareable data provides new utility for high-performance and data-center scale systems.
Existing parallel programming frameworks and APIs, such as OpenSHMEM \cite{chapman2010introducing} or MPI \cite{gropp1996high}, can be ported to the Zeno architecture.
Software overhead related to address translation and the networking stack can be reduced by exploiting the hardware support for globally shared Namespace data.
The standalone address space abstraction offered by Namespaces makes it possible to map all of the storage media (hard disks, flash, etc.) in a system into the memory address space.
With the proper system support, storage media becomes directly accessible to software with loads and stores, reducing filesystem overhead and file I/O access latency.
The shareable and hierarchical nature of Namespaces allow filesystem management software to derive smaller, per-file Namespaces for other software to access.
Efficiently splitting large, irregular, in-memory datasets, such as graphs, across nodes is non-trivial.
The addresses of existing systems are associated with a physical node, forcing programmers to carefully consider the data placement for each node.
With the Namespace address space abstraction decoupled from the physical memory of each node, programmers no longer need to manually split their data between nodes.
Instead, the Zeno architecture can leverage remote data caching to improve data locality based on application access patterns.

Our contributions in this work include:
\begin{itemize}
\item The introduction, design, and implementation of Zeno - a new Namespace-based capability architecture for large-scale systems. 
\item An evaluation of Zeno's scalability and performance overhead as a multi-node system.
\item A validation of the Zeno architecture through a hardware/FPGA implementation to evaluate area and power overheads, and programming complexity.
\item Zeno provides an encouraging solution to the memory safe and security problems found on many multi-core and multi-node processor systems. 
\end{itemize}

\section{Related Work}

Many previous works have explored different capability model implementations to address memory safety and security.
The CHEx86 architecture implements capabilities at the microcode level, enabling support for legacy binaries \cite{sharifi2020chex86}. 
The Intel Memory Protection Extension (MPX) is an ISA extension providing bounds checking on pointers \cite{oleksenko2018intel}. 
Intel MPX uses the address of the pointer to compute the bound's memory address, allowing bounds to be located without any additional metadata.
Another architecture named CHERI achieves pointer safety by replacing traditional pointers with capability pointers encoded with compressed bounds metadata \cite{woodruff2019cheri} \cite{davis2019cheriabi}. Temporal safety is achieved by periodically sweeping memory to clean up stale capabilities \cite{xia2019cherivoke}.

Existing capability architectures have not extended their capability models to multi-node systems.
Certain features, such as sweeps to reclaim physical memory, will not scale up for large systems.
Challenges related to multi-node shared memory and remote memory access are unaddressed by existing works.

\section{Security Properties}

The Zeno architecture aims to support safe and secure memory in large scale systems.
In addition to contending with spatial and temporal memory safety challenges faced by other systems, multi-node systems must consider the risks of remote memory access.
When multiple compute nodes share data and act as a single system, a single corrupt node may manipulate the memory of many other nodes, corrupting the entire system.
Specific memory safety and security challenges for large-scale systems include:

\begin{enumerate}
    \item Spatial memory safety - Memory reads and writes beyond the scope of the intended memory buffer can corrupt memory and alter the execution flow of programs.
    \item Temporal memory safety - Use-after-free and double free errors in programs can lead to memory corruption and unexpected execution flows.
    \item Remote memory access - In a system with remote memory access support, malicious software executing on a node can abuse that support to make unauthorized requests for data on other nodes. If malicious software is able to elevate its privileges, software authentication may be bypassed.
    \item Least privileges in globally shareable memory - As programs begin or end, and as program phases change, the data each node must access also changes.
    Access to globally shareable data must be revocable to ensure each node may only access the data necessary for its computation.
    Shared access to data increases the attack surface of a program, creating more opportunities to corrupt data or alter the program's execution flow.
\end{enumerate}

Zeno addresses these challenges with its Namespace capability model.
Spatial memory safety is maintained with byte-granular minimum/maximum access bounds on each Namespace capability.
Bounds are set when the Namespace is created and cannot be modified afterwards.
Bounds of derived Namespaces are monotonically decreasing; A child Namespace cannot have larger bounds than the parent Namespace it was derived 
from.

Namespace revocation supports temporal memory safety by invalidating the access permission metadata associated with the given Namespace capability. 
Valid Namespace Capabilities may remain in memory or on-chip after revocation, but attempts to access the Namespace data with them will fault.
Use-after-free or double-free programming errors cannot corrupt system memory when the Namespace of the freed memory has been revoked. 
Invalidating Namespace metadata means the physical memory previously used by the Namespace address space abstraction is immediately available for re-use in another Namespace.
However, Namespace Capabilities in memory or on-chip must be swept for and invalidated before the Namespace ID is reused.
Using Namespace IDs with as many bits as addresses
Using many bits in the Namespace ID, e.g. 64-bits, ensures that plenty of Namespace IDs are available for use while the system sweeps for revoked Namespace IDs.

A system with remote memory access support must validate access requests received from the system interconnect.
Software based validation of requests carries a high performance overhead and adds complexity to the system software stack, increasing the opportunities for exploits.
The Zeno architecture's capability model extends beyond a single node, to consider multi-node systems.
Namespaces share a common representation for Namespace IDs and access permissions across every node, enabling all access validation to be performed in hardware.
Each node accesses the same shared Namespace metadata storage to create a single system level view of the current Namespace access permissions.
Enforcing the  the same access permissions across the system at the hardware level minimizes the additional attack surface created by remote memory access.

Zeno Namespace capabilities are hierarchical, supporting the concept of least privilege.
A child Namespace can be derived from a parent Namespace with reduced permissions for sharing.
Both the child and parent Namespace reference the same address space abstraction, but with different accessible ranges.
The child Namespace may not have access to all of the data a parent Namespace does.
Revoking a Namespace also revokes all of the children of that Namespace, preventing continued access to Namespace data through another child Namespace.
\section{The Zeno Architecture}
\label{arch}

\subsection{Zeno Architecture Overview}

\begin{figure}[t]
    \centering
     \includegraphics[width=0.99\columnwidth]{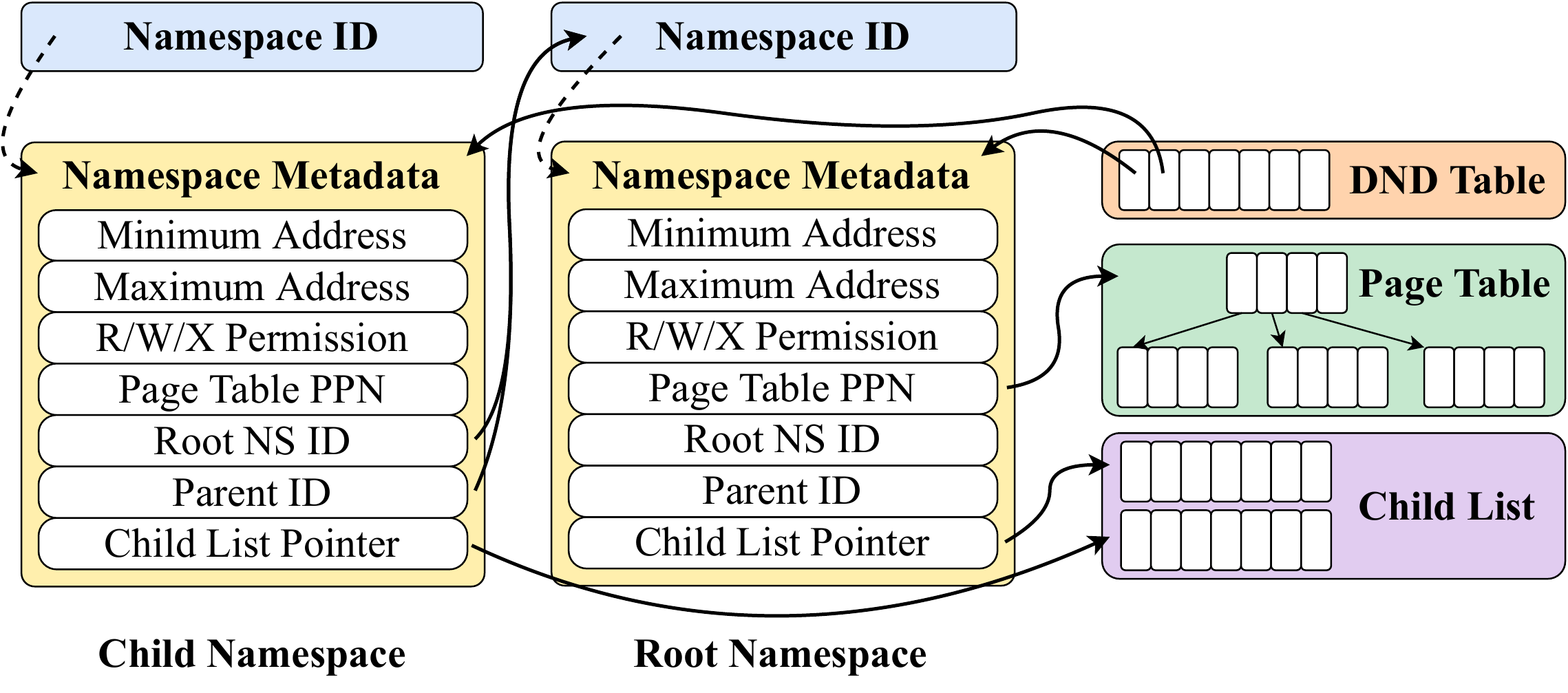}
     \vspace{-.3in}
     \caption{Namespace metadata format.}
     \label{fig:ns_metadata}
\end{figure}

The Zeno architecture uses an extended addressing model \cite{xbgas} to support the large addresses and the use of multiple Namespace address space abstractions simultaneously.
Extended portions of addresses encode a Namespace ID to reference Namespace Metadata and identify which address space abstraction the base address references.
Figure~\ref{fig:ns_metadata} shows the contents and layout of Namespace metadata.
Namespace Metadata stores access permissions, in addition to information about the hierarchical organization of Namespaces.
Minimum and maximum address bounds describe the Namespace's accessible range of addresses in the address space abstraction.
Read, write, execute, and valid permission bits describe the allowable memory operations.
A physical page number of the Namespace page table is also included in the metadata.
Each parent and child Namespace provides access to the same address space abstraction and therefore shares a page table.
Metadata includes the Namespace's Root Namespace ID, that is, the Namespace ID at the root of this hierarchy of Namespaces. A Root Namespace has no Parent Namespace and is useful for building address space IDs in the microarchitecture.
Many Root Namespaces will exist simultaneously in a Zeno system.
Though no current operations require it, the direct Parent Namespace ID is also stored in the Namespace Metadata to support system debugging and future operations that may reason about the Namespace hierarchy.
A pointer to a list of children Namespaces is included in the metadata of each Namespace to support recursive Namespace Revocation.
All Namespace Metadata is stored in the Distributed Namespace Directory, a dedicated Namespace accessible only to hardware and trusted firmware responsible for managing Namespace data.

The Namespace ID acts as an access token, granting access to Namespace Data.
Namespace IDs are created by hardware for the requesting software context.
Software is free to share its available Namespace IDs with other software executing anywhere in a multi-node system.
Each node in the system views a shared Namespace as the same address space abstraction.
Without the Namespace ID access token, software cannot access a Namespace.
Namespace IDs are protected by tag bits in memory and on the processor die to prevent forgeries.
The hardware-enforced address bounds of Namespace capabilities provide spatial memory safety.
Namespace revocation supports temporal memory safety by removing all access permissions granted by a Namespace without the need to invalidate the Namespace ID access token everywhere in the system.

%%%%%%%%%%%%%%%%%%%%%%%%%%%%%%%%%%%%%%%%%%%%%%%%%%%%%%%%%%%%%%%%%%%%%%%%%%%%%%%%%%%%%%%%%%%%%
\subsection{Zeno Core Level Description}
%%%%%%%%%%%%%%%%%%%%%%%%%%%%%%%%%%%%%%%%%%%%%%%%%%%%%%%%%%%%%%%%%%%%%%%%%%%%%%%%%%%%%%%%%%%%%

\begin{figure}[t]
    \centering
    \includegraphics[width=0.85\columnwidth]{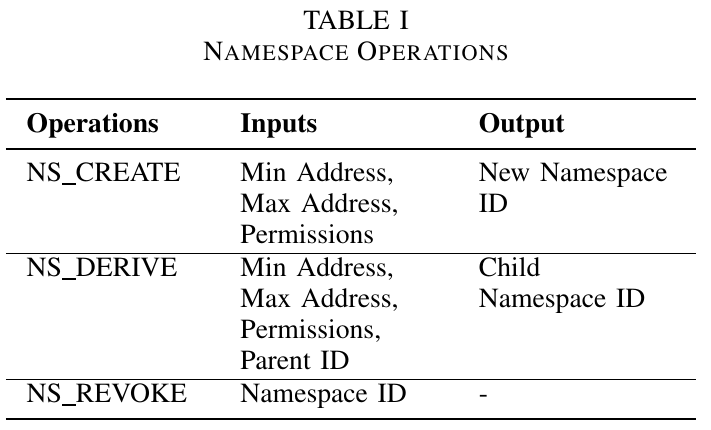}
    \label{fig:table1}
    \vspace{-0.2in}
\end{figure}

\begin{figure*}[t!]
    \centering

     \vspace{0.1in}
     \includegraphics[width=0.90\textwidth]{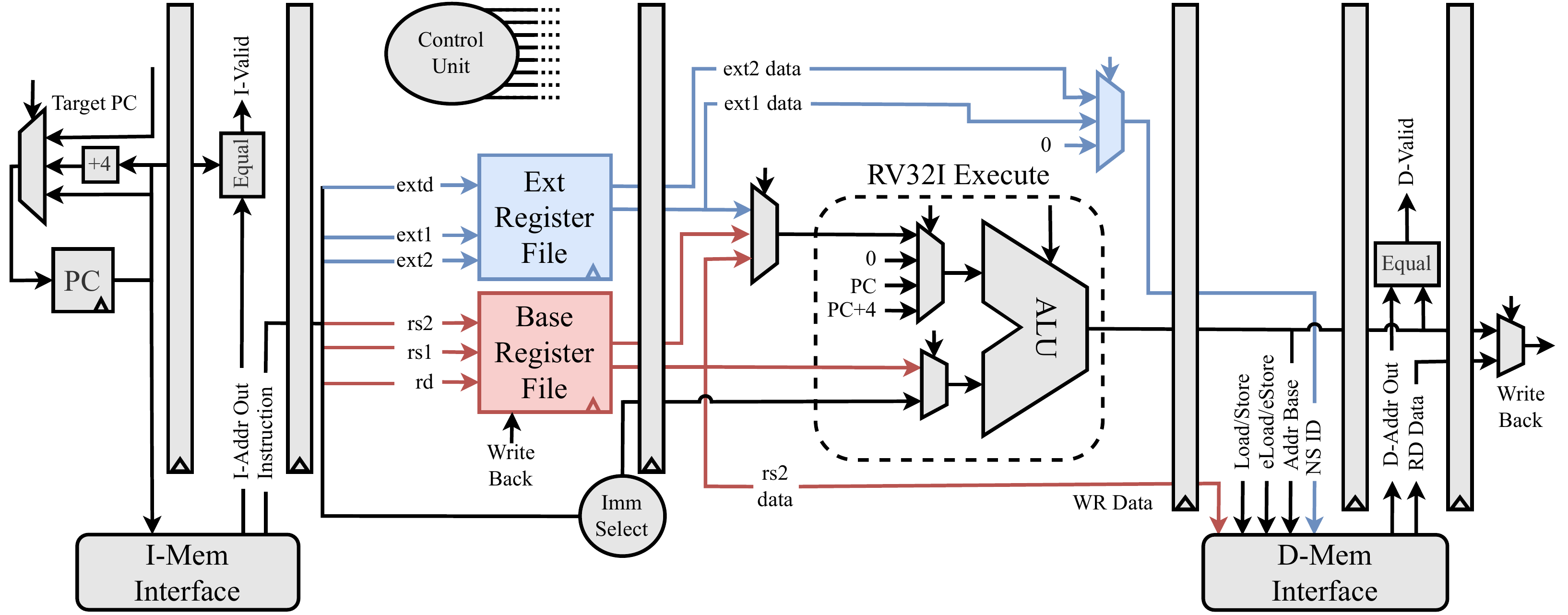}
     \vspace{-0.2in}
     \caption{Zeno 7-stage pipeline micro-architecture.}
     \label{zeno_core}

    \vspace{-0.1in}
\end{figure*}

Zeno cores include an extended set of 64-bit registers to store Namespace IDs.
Figure~\ref{zeno_core} shows the micro-architecture of a Zeno core.
Signals and modules in blue depict logic added to a baseline RISC-V core to support Namespace IDs.
Signals and modules in red depict the base rv64i signals.

Each core in a Zeno system must support Namespace creation, derivation, and revocation.
Table I describes the inputs and outputs of each operation.
In this work, the Namespace operations are each encoded into a single instruction to ensure they happen atomically.
The extended register file is used to store Namespace IDs and pack up to four register reads into a single instruction while conforming to RISC-V's standard encoding for register selection.

The \texttt{NS\_CREATE} instruction sets the Namespace metadata in memory and writes a valid Namespace ID into the extended register file. 
To minimize changes to the processor datapath, the \texttt{NS\_CREATE} instruction is decoded into several RISC operations to store appropriate metadata values in memory.
The \texttt{NS\_DERIVE} instruction is similar to the \texttt{NS\_CREATE}, but also requires the Namespace ID of the Parent Namespace.
The \texttt{NS\_REVOKE} instruction requires only the Namespace ID to be revoked.
If a Namespace with valid children is revoked, the system must recursively revoke them with either hardware/microcode operations, or a fault and trusted firmware.

%%%%%%%%%%%%%%%%%%%%%%%%%%%%%%%%%%%%%%%%%%%%%%%%%%%%%%%%%%%%%%%%%%%%%%%%%%%%%%%%%%%%%%%%%%%%%
\subsection{Zeno Node Level Description}
%%%%%%%%%%%%%%%%%%%%%%%%%%%%%%%%%%%%%%%%%%%%%%%%%%%%%%%%%%%%%%%%%%%%%%%%%%%%%%%%%%%%%%%%%%%%%
\label{sec:node_level_description}

The Zeno architecture does not require any changes to on-chip instruction/data caches, but does require additional Memory Management Unit (MMU) logic.
To improve performance, Zeno includes an on-chip Metadata Cache inside the MMU to quickly access Namespace Metadata.
When a memory access is issued by the Zeno core, the Metadata Cache uses the Namespace ID to search the cache for the associated Metadata in parallel with TLB lookup and the virtual address translation.
On a hit, the metadata is read out to the permissions checking logic, which validates the requests and allows the memory request to continue.
On a miss, the metadata cache can either raise a fault or fetch the metadata with logic similar to a page table walker. 
These options mirror the RISC-V address translation ISA specification.

In the MMU, the Root Namespace IDs are added to the TLB tags to distinguish between the different Namespace address space abstractions.
We denote this TLB the Namespace-TLB (N-TLB).
The page table walker is updated to use the Namespace Metadata page table physical page number rather than RISC-V's SATP register.

Remote memory is accessed through a remote-memory cache stored in local DRAM, 
allowing remote data to appear local to software and address translation mechanisms.
The remote data cache scheme allows hardware to fetch the remote data without any intervention from local software or remote CPUs, simplifying programming and improving performance.
For scalability, remote memory cache coherence is managed in software.
To support hardware fetching of remote data, each Zeno node includes a network interface for remote memory access.
The network interface sends Namespace capability memory requests to remote nodes and serves request received from other nodes.
The network interface is coherent with the Node's local cache hierarchy to maintain coherence within a node.
The network interface must perform the same Namespace permissions checks as the MMU connected to each core and VIPT L1 cache. 
A strait-forward network interface implementation re-uses the cache hierarchy typically connected to a core and connects it to a DMA engine serving requests across the system interconnect.

\begin{figure}[t!]
    \centering

     \vspace{0.1in}
    \includegraphics[width=0.75\columnwidth]{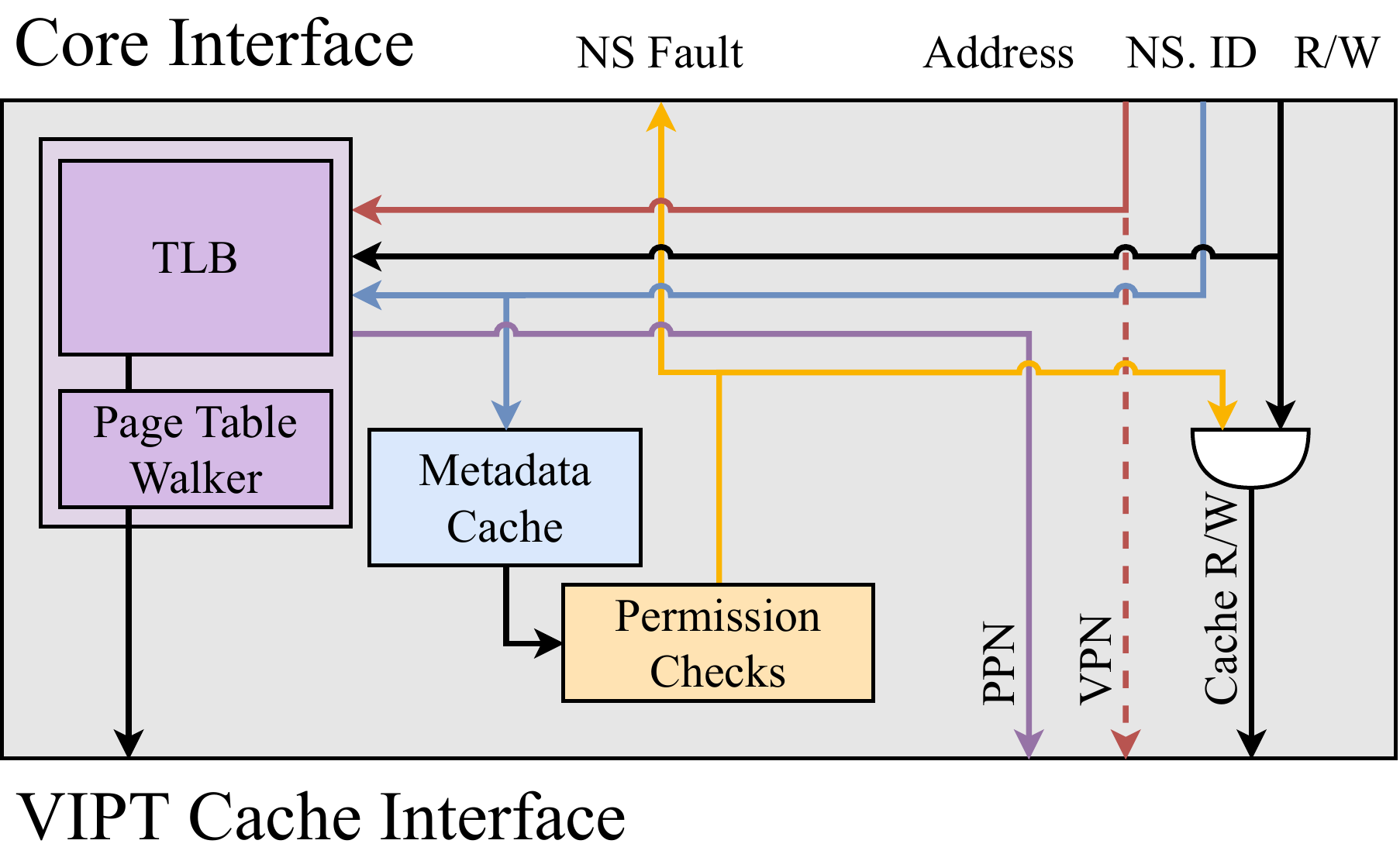}
    \vspace{-0.1in}
    \caption{Zeno node architecture.}
    \vspace{0.2in}
    \label{zeno_node}
    
    \includegraphics[width=0.99\columnwidth]{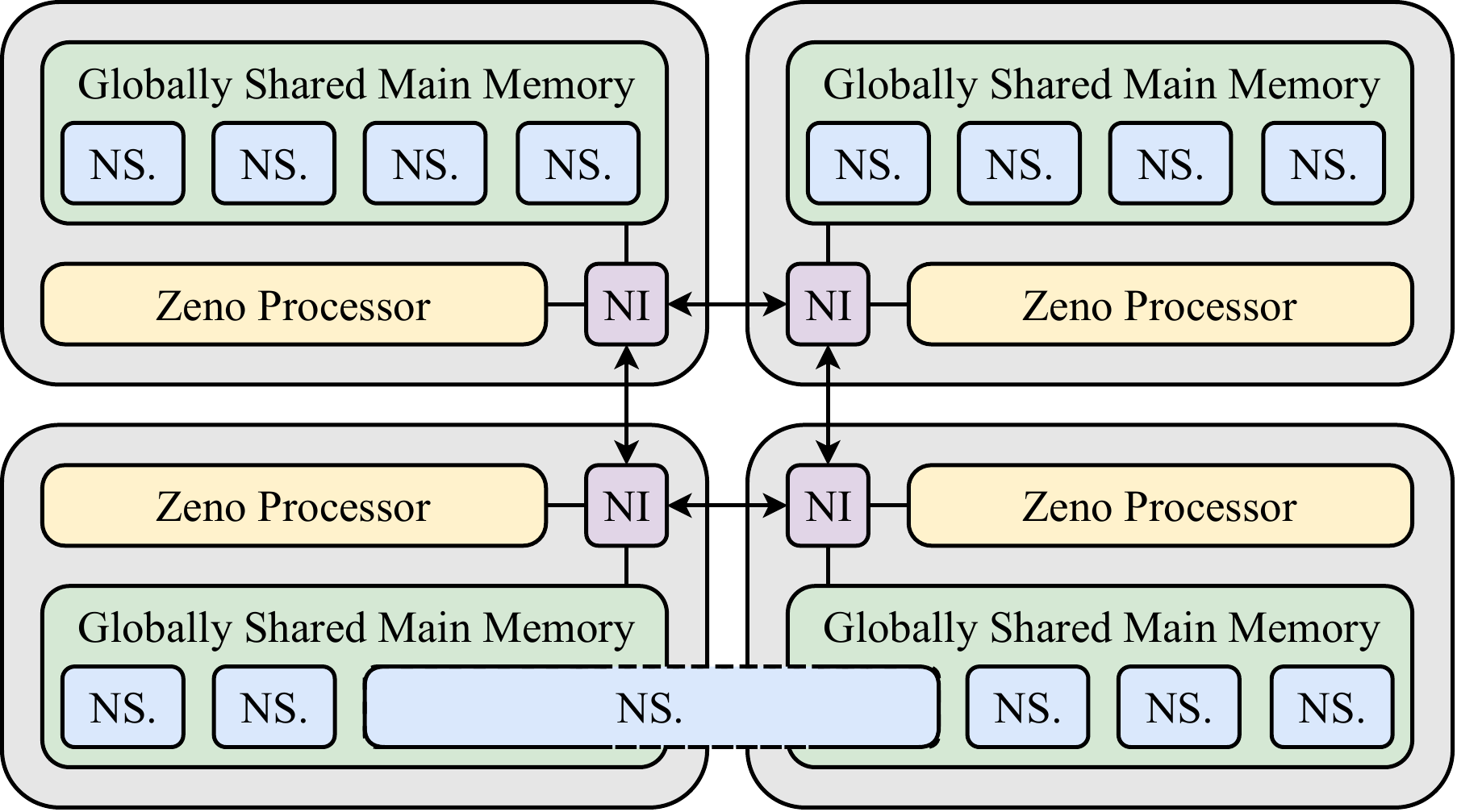}
    \vspace{-0.2in}
    \caption{A four node Zeno system}
    % \vspace{0.1in}
    \label{system}

    \vspace{-0.1in}
\end{figure}

%%%%%%%%%%%%%%%%%%%%%%%%%%%%%%%%%%%%%%%%%%%%%%%%%%%%%%%%%%%%%%%%%%%%%%%%%%%%%%%%%%%%%%%%%%%%%
\subsection{Zeno System Level Description}
%%%%%%%%%%%%%%%%%%%%%%%%%%%%%%%%%%%%%%%%%%%%%%%%%%%%%%%%%%%%%%%%%%%%%%%%%%%%%%%%%%%%%%%%%%%%%
Zeno connects distributed nodes of compute resources together into a single system.
Example node configurations include: a processor and memory, just memory resources, or I/O disk resources.
Figure~\ref{system} shows an example four node Zeno system.
Each node has a multi-core processor and globally accessible main memory.
Zeno is intended to support large scale systems, i.e. many racks worth of compute and storage nodes.
However, smaller rack-scale and single node systems can also be implemented.
Zeno does not require a specific interconnect topology or protocol to connect multiple nodes together, as long as they are connected through  Namespace validating Network Interfaces.

\section{Experimental Setup}

To measure the overhead and performance of the Zeno architecture, we developed an ISA simulator in C++.
% Simulation parameters
%Simulation parameters were selected to model our FPGA implementation.
The simulator models an in-order core with one instruction issued per cycle.
Connected to the local cache hierarchy are two  L1 instruction and data caches and a shared L2 cache.
The system interconnect is modeled as a 2D mesh of routers. Router latency is based on our FPGA implementation.
Table II presents the access latency for different system components.

A file with RISC-V rv64i and Zeno assembly instructions is read in by the simulator for execution.
The simulator executes the assembly instructions and models the system's register files, memory hierarchies, and performance counters.
Performance counter values are output at the end of program execution.
The performance counters track the number of instructions committed, the total number of cycles in program execution, as well as N-TLB and metadata cache hits and misses.
The CPU cycle counter tracks the number of clock cycles the CPU is not stalled and waiting on a memory request.
The NLB counter tracks the time spent looking up values in the NLB's N-TLB and metadata cache.
Local and global memory access counters track the runtime spent waiting on the memory operations to complete.
The link latency and routing delays of the system's NoC and off-chip network is factored into global memory access latency.
Network delays are modeled on the distribution of latency in the Zeno RTL implementation which extends the BRISC-V Platform \cite{bandara2019briscv}.

Three benchmarks are run on the simulator:
1) a  ``Get Transfer'' benchmark that makes 4kB accesses to one Namespace on each node in the system,
2) a ``Random Memory Access'' benchmark that randomly accesses 4B of data in an array of 128 32kB Namespaces evenly spread across the system, and
3) a parallel ``Integer Sort`` application that implements the bucket sort algorithm to sort 64k 4B integers using one Namespace per bucket.

\section{Results}

\begin{figure}[t]
    \centering
    \includegraphics[width=0.95\columnwidth]{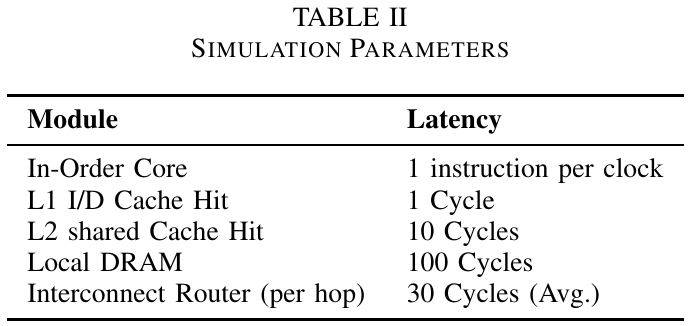}
    \label{fig:table2}
\end{figure}

\begin{figure}[t!]
	\centering
% 	\vspace{-0.1in}
        \includegraphics[width=0.97\columnwidth]{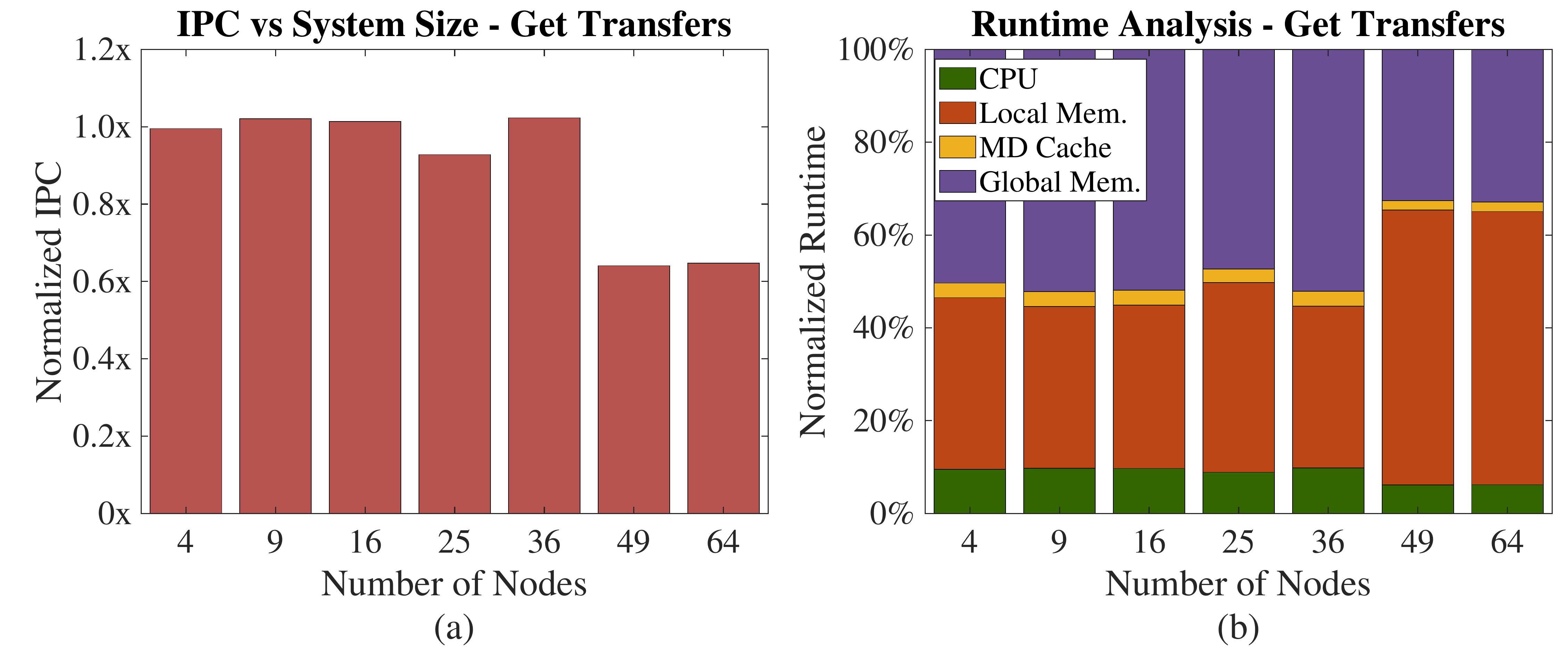}\vspace{-0.05in}
        \vspace{-0.075in}
        \caption{Get Transfers IPC and runtime analysis.}
        \label{get_both}
        \vspace{.1in}
        \includegraphics[width=0.97\columnwidth]{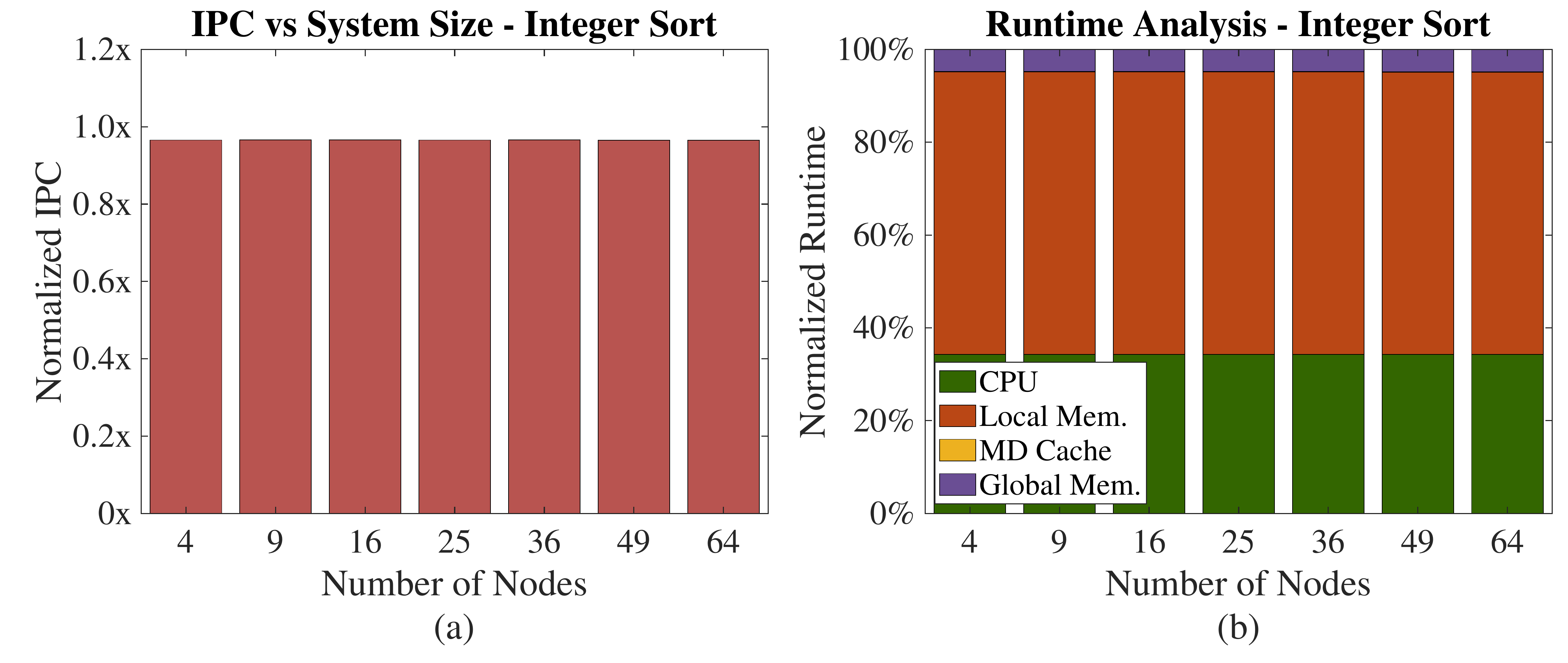}\vspace{-0.1in}
        \vspace{-0.075in}
        \caption{Integer Sort IPC and runtime analysis.}
        \label{sort_both}
 	\vspace{-0.2in}
\end{figure}

\begin{figure*}[t!]
	\centering
% 	\vspace{-0.1in}
	\begin{minipage}{\columnwidth}
		\centering
        \includegraphics[width=0.9\textwidth]{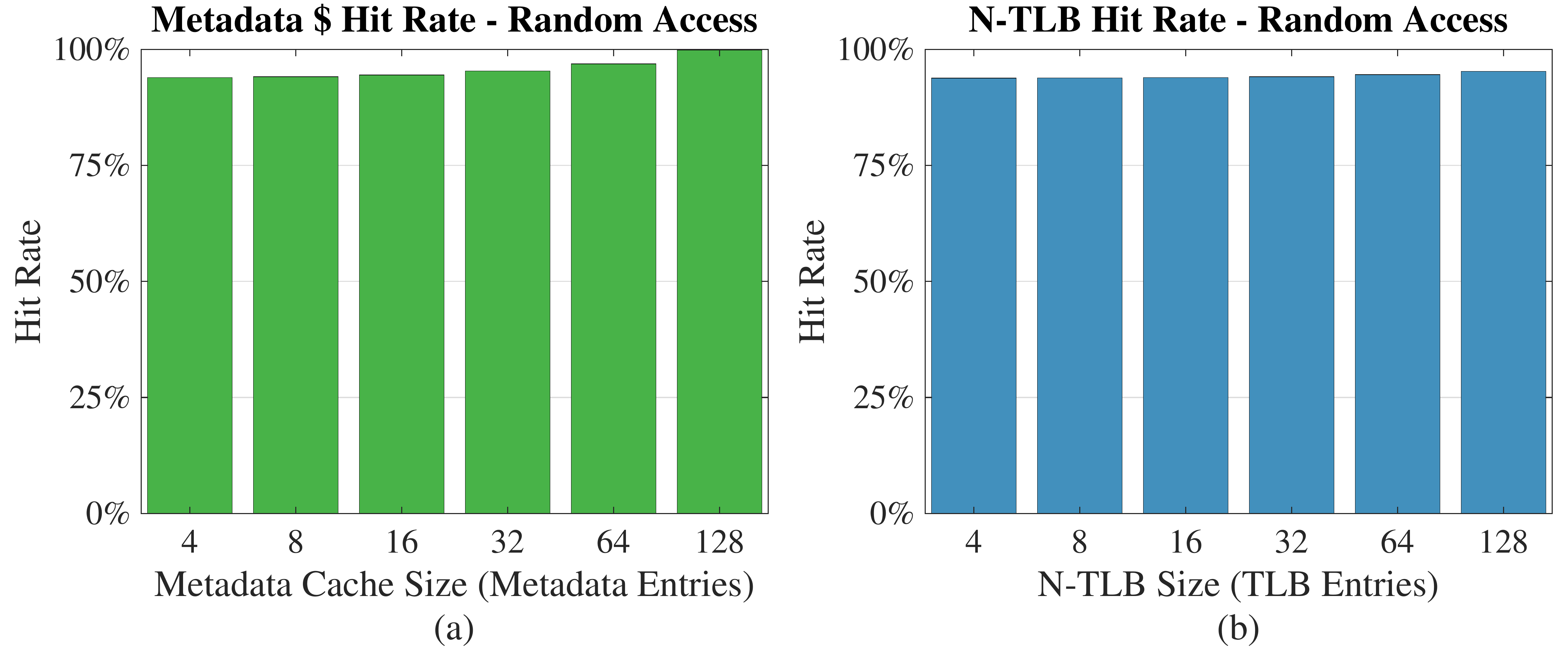}\vspace{-0.05in}
        \vspace{-0.075in}
        \caption{Random memory access  hit rates. }
        \label{random_hit_rate}
	\end{minipage}
	\begin{minipage}{\columnwidth}
		\centering
        \includegraphics[width=0.9\textwidth]{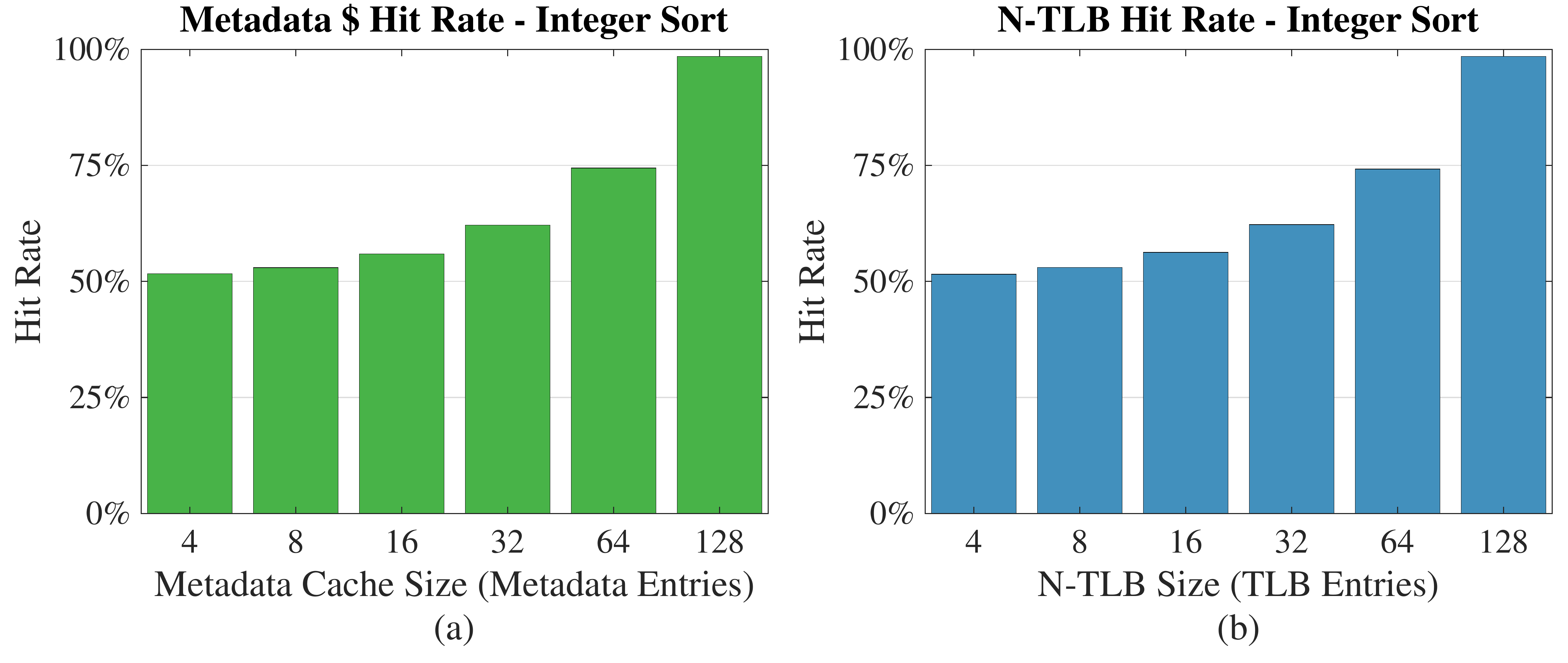}\vspace{-0.1in}
        \vspace{-0.075in}
        \caption{Integer sort  hit rates.}
        \label{sort_hit_rate}
	\end{minipage}
 	%\vspace{-0.1in}
\end{figure*}

\begin{figure*}[t!]
	\centering
% 	\vspace{-0.1in}
	\begin{minipage}{\columnwidth}
		\centering
        \includegraphics[width=0.9\textwidth]{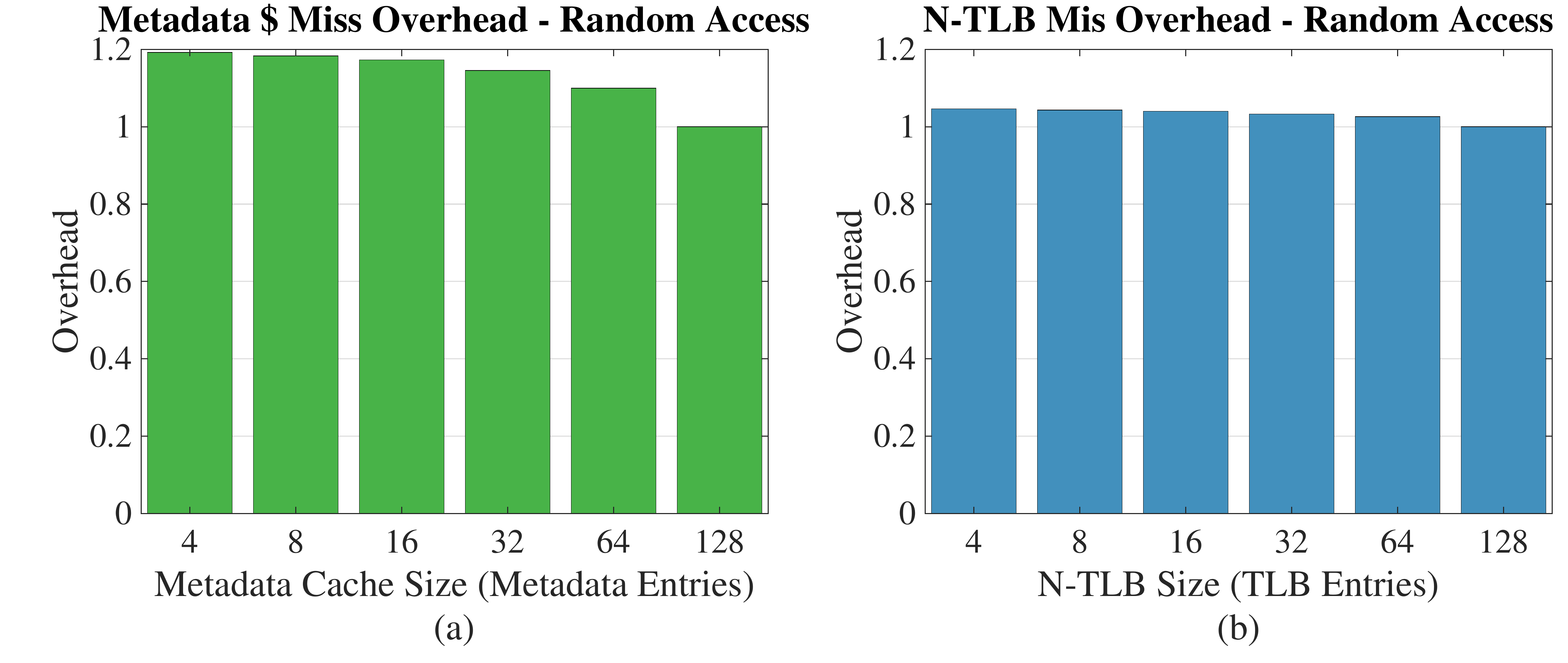}\vspace{-0.05in}
        \vspace{-0.075in}

        \caption{Random Memory Access overheads. }
        \label{random_NLB_overhead}
	\end{minipage}
	\begin{minipage}{\columnwidth}
		\centering
        \includegraphics[width=0.9\textwidth]{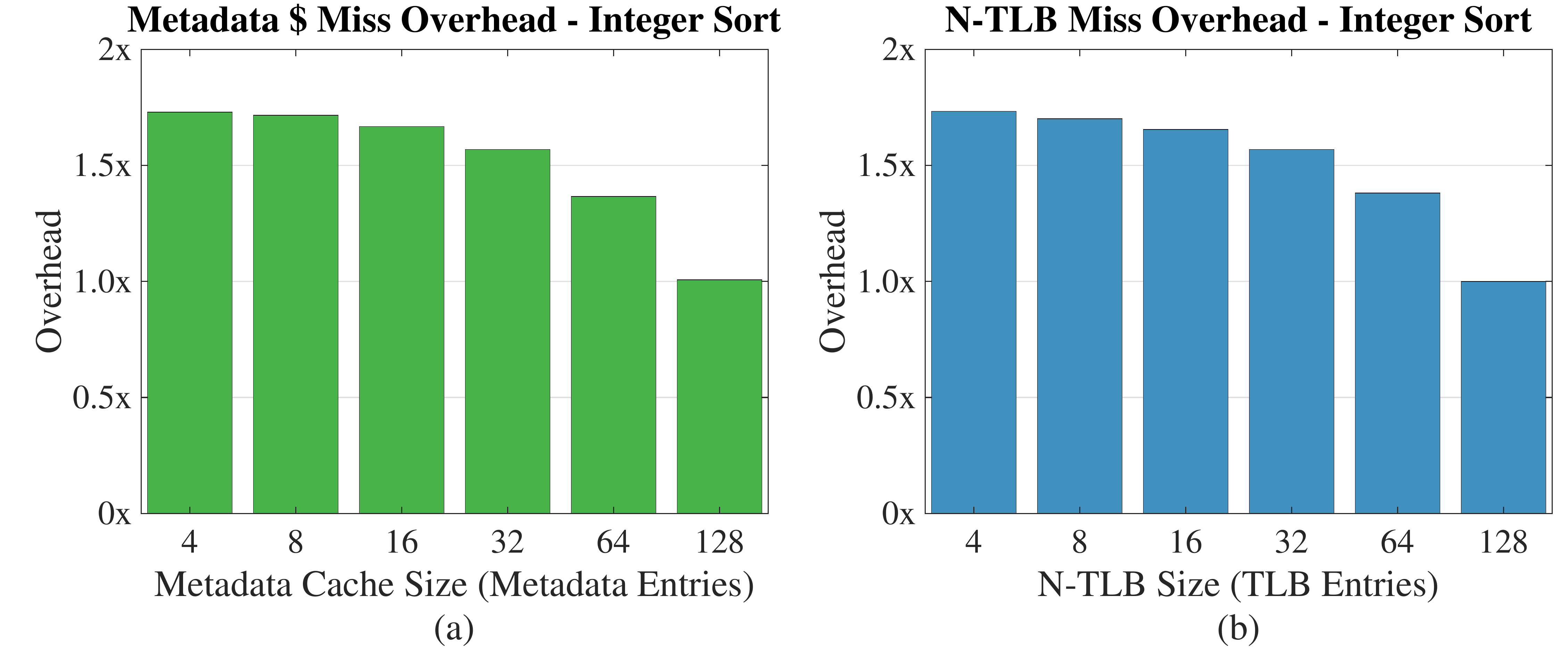}\vspace{-0.1in}
        \vspace{-0.075in}
        \caption{Integer Sort overheads.}
        \label{sort_NLB_overhead}
	\end{minipage}
   	\vspace{-0.1in}
\end{figure*}

\begin{figure*}[t]
	\centering
	%\vspace{-0.15in}
	\begin{minipage}{\columnwidth}
		\centering
        \includegraphics[width=0.85\textwidth]{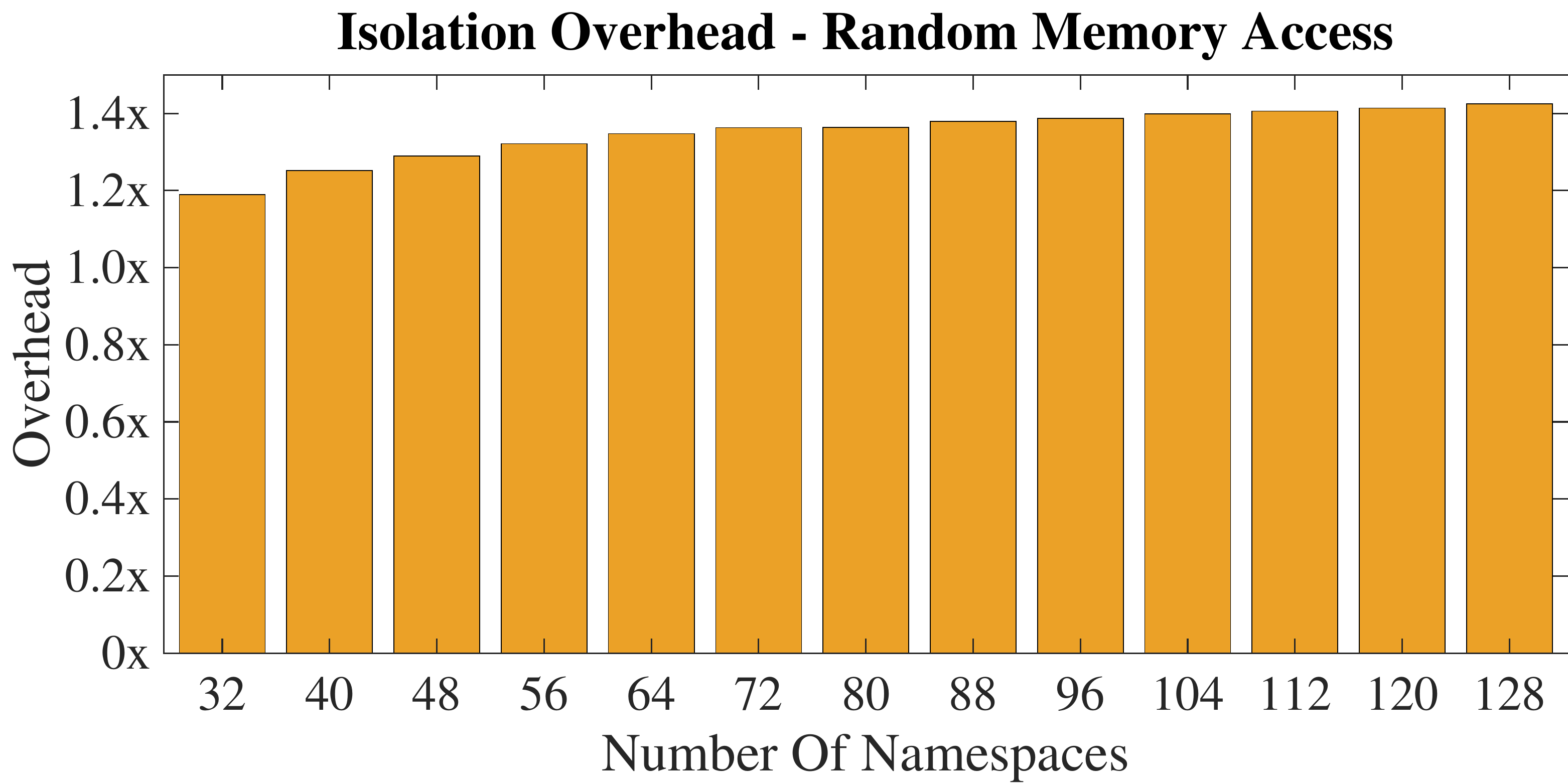}\vspace{-0.05in}
        \vspace{-0.05in}
        \caption{Random Access global memory overhead.}
        \label{random_overhead}
	\end{minipage}
	\begin{minipage}{\columnwidth}
		\centering
        \includegraphics[width=0.85\textwidth]{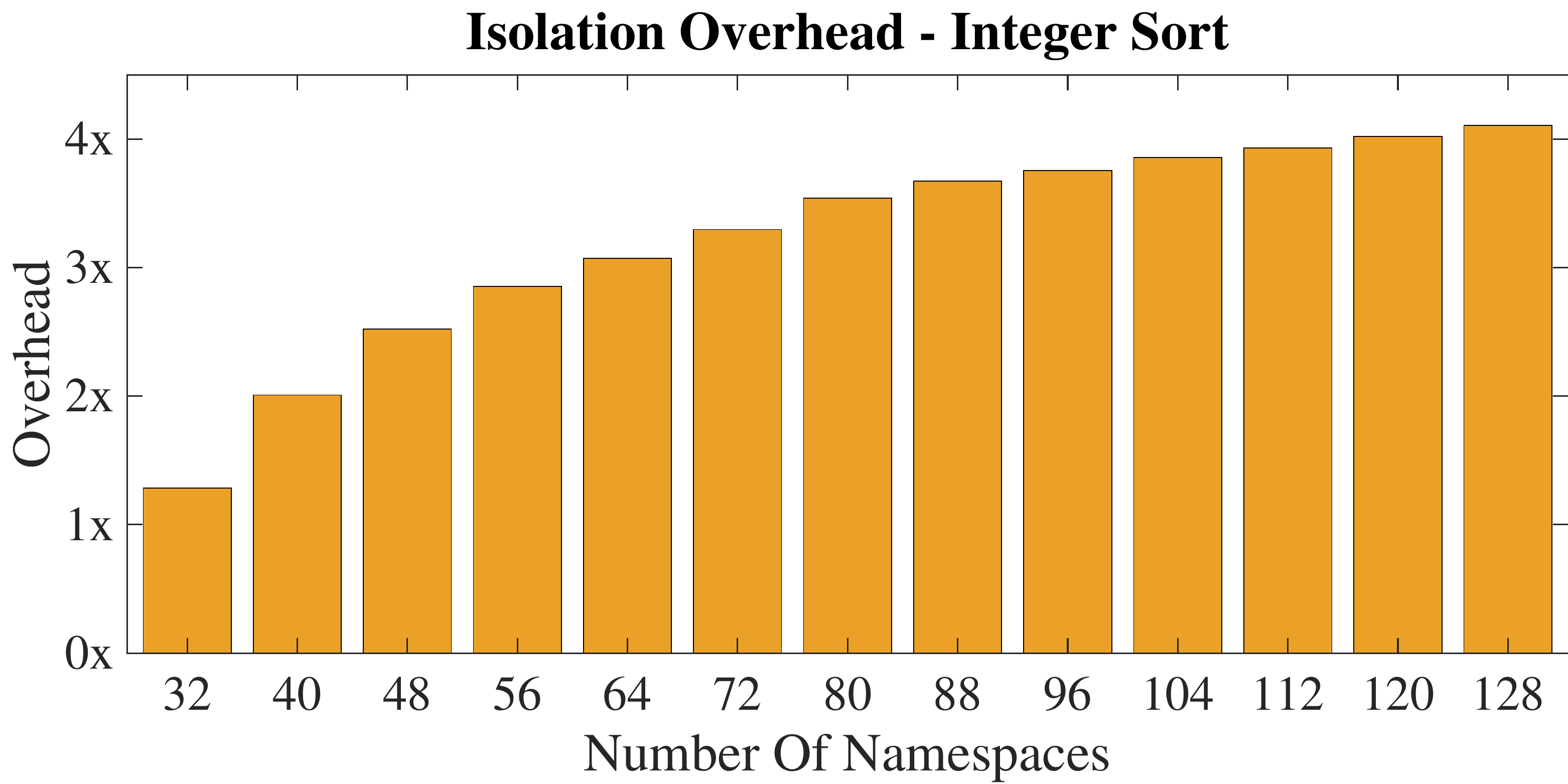}\vspace{-0.1in}
        \vspace{-0.05in}
        \caption{Integer Sort global memory overhead.}
        \label{sort_overhead}
	\end{minipage}
	\vspace{-0.15in}
\end{figure*}

\subsection{Simulation Results}

Figures~\ref{get_both}(a) and \ref{sort_both}(a) demonstrate the performance scaling of the Get Transfer and Integer Sort benchmarks on Zeno architecture as more nodes are added to the system's global memory space.
Figures~\ref{get_both}(b) and \ref{sort_both}(b) show the fraction of runtime spent constrained by each major subsystem in the architecture (the CPU, NLB, local memory hierarchy, and global memory hierarchy) for each benchmark.
Both the Get Transfers and Integer Sort benchmarks are each simulated in a Zeno configuration with only 32 N-TLB and metadata cache entries to ensure the benchmarks working set could not be completely buffered in either structure.
Systems are arranged in 2D meshes with sizes ranging from 2x2 to 8x8.
IPC results are normalized to the IPC of a baseline 2x2 xBGAS \cite{xbgas} system with none of Zeno's Namespace capability memory protections.

Figure~\ref{get_both}(a) indicates that the performance of the Get Transfers benchmark remains steady for 4-36 node systems and decreases in the 49 and 64 node systems.
To investigate the cause of this performance reduction, we plot the fraction of runtime dedicated to CPU execution, local and global memory accesses, as well as NLB lookups.
Note that while the Get Transfers benchmark sees a reduced IPC for larger systems, this is not caused by an increased amount of global memory accesses or NLB lookups.
Instead the additional latency stems from extra local memory accesses as the working set of the benchmark increases with the size of the system.

Figure~\ref{sort_both}(a) shows that the Integer Sort benchmark scales nearly perfectly with an increasing system size.
This is because remote Namespaces are only accessed to set up the parallel bucket sort and collect the results. 
Figure~\ref{sort_both}(b) demonstrates that, although the Integer Sort runtime is dominated by CPU execution cycles and local memory accesses, the fraction of runtime spent on global memory accesses remains steady. 

Taken together, Figures~\ref{get_both}  and \ref{sort_both} demonstrate that the Namespace based access scheme forms a scalable system that will not be limited by the increased overhead of Namespace Metadata lookups.
For the Get Transfers benchmark, the Metadata Cache lookup overhead represents 2-3\% of the total program runtime.
For the Integer sort benchmark, the Metadata Cache lookup overhead represents less than 0.1\% of the total program runtime.

Next we examine the design trade-off between different MMU configurations with the Random Memory Access and Integer Sort benchmarks.
Including both a cache for metadata and a Translation Lookaside Buffer creates a rich design space where it may be difficult to find an optimal size for each structure.
A larger Metadata Cache and N-TLB will lead to greater performance but will also increase power and area requirements.
Furthermore, it is not clear how a fixed area should be shared between the N-TLB and the metadata cache in the new Zeno architecture.

To explore the NLB design space, we simulate both benchmarks with metadata cache and N-TLB sizes of powers of two ranging from 4-128.
This is six sizes per structure or 36 simulations per benchmark.
Figure~\ref{random_hit_rate}(a) and \ref{sort_hit_rate}(a) depicts the Metadata Cache hit rates with a 32 entry N-TLB.
Figure~\ref{random_hit_rate}(b) and \ref{sort_hit_rate}(b) depicts the N-TLB hit rates with a 32 entry Metadata Cache.
A 16 node Zeno system is used in each of the simulations.
Each benchmark uses 128 Namespaces.
The trends in hit rates for increasing metadata cache and N-TLB sizes are similar.
The hit-rates suggest that neither the metadata cache or N-TLB can dominate the MMU area, as performance steadily improves in both benchmarks with larger structures.

To gain more insight, we plot the overhead of additional global memory accesses in Figures~\ref{random_NLB_overhead} and \ref{sort_NLB_overhead} for the same benchmarks and MMU configurations used in Figures~\ref{random_hit_rate} and \ref{sort_hit_rate}.
The reduction in overhead in the Integer Sort benchmark as the NLB structures increase in size (Figure~\ref{sort_NLB_overhead}(a) and (b)) is similar, indicating that the metadata cache and N-TLB are equally important in the performance of the Integer Sort benchmark. 
However, comparing Sub-figures (a) and (b) in Figure~\ref{random_NLB_overhead} shows that the Random Memory Access benchmark benefits more from a large metadata cache than a large N-TLB.
The better performance of the large metadata cache stems from the greater fraction of addressable memory covered by a full Metadata Cache than an N-TLB of the same size. 
For applications with better spatial or temporal locality, this difference is less pronounced.
In the random memory accesses used here (but also in HPC applications such as graph processing) the impact of this improved global memory space coverage becomes apparent.

Finally, we examine the performance overheads of programming with many or few Namespaces.
Using many Namespaces enables greater isolation and increased memory safety but at the expense of performance, as more Metadata Cache and N-TLB misses are inevitable.
Figures~\ref{random_overhead} and \ref{sort_overhead} quantifies this trade-off by presenting the overhead of increased global memory accesses caused by metadata cache and N-TLB misses.
Again, simulations are run on a Zeno system with 16 nodes where each NLB includes a 32 entry metadata cache and 32 entry N-TLB.
The additional cycles spent accessing global memory are normalized against a plain xBGAS system that implements a flat 128-bit memory space without any security protections.
The smallest test limits the number of Namespaces to the size of the metadata cache and practically eliminates the overhead of fetching Namespace Metadata. The initial 32 metadata misses are insignificant compared to the total program runtime.

Figure~\ref{random_overhead} shows that increasing the number of Namespaces from 32 to 128 in the Random Memory Access benchmark leads to a 1.4x overhead.
The overhead of going from 32 to 128 Namespaces in the Integer Sort benchmark, shown in Figure~\ref{sort_overhead}, is higher at 4.1x.
The relatively low overhead of the Random Memory Access benchmark is likely because of the limited effectiveness of the Metadata Cache and N-TLB with a random access pattern in the first place.
Since there was little spatial or temporal locality for the caches to exploit at 32 Namespaces, frequent global memory accesses were required to fetch Namespace Metadata and address translations from memory.
Utilizing 128 Namespaces instead of 32 did not significantly worsen any locality in the benchmark since there was little to begin with.
The Integer Sort benchmark has more spatial locality to exploit, especially at the end of the computation when results are collected on a single node.
Figure~\ref{sort_overhead} shows that the overhead of additional Namespaces can become more significant.
However, the additional overhead begins to level off as more Namespaces are included in the program,
indicating that  spatial or temporal locality  in the program allows the Metadata Caches and N-TLB to remain effective.

\vspace{-0.05in}
\subsection{Synthesis Results}

Table III presents synthesis results for the Zeno architecture, an xBGAS enabled processor without Zeno's additional micro-architecture features and a baseline RISC-V system.
Each system includes includes a seven stage pipelined rv64ima core with 32kB, 8-way instruction and data L1 caches, a unified 256kB 8-way L2 cache and Memory Management Unit. 
The xBGAS-enabled system and the Zeno architecture extend the baseline RISC-V core with support for the xBGAS ISA extension.
Additional micro-architecture features of the Zeno architecture include the Metadata Cache, modified TLB and permission checking logic.
The Zeno system is configured with a 1024 entry, 8-way set associative N-TLB and a 128-entry, 8-way set associative Metadata Cache.
The targeted FPGA is an Altera Stratix V (5SGXEA7N2F45C2) \cite{altera2019stratix}.
Quartus II 15.0 is used for synthesis.
Area results for Adaptive Logic Modules (ALMs) \cite{altera2019stratix}, registers and BRAM bits are reported.
The percentage of FPGA ALMs and BRAMs used is reported in parenthesis next to the raw number of resources used.

Most of the resource overhead necessary to implement the xBGAS ISA extension is dedicated to the additional register file, with instruction decode logic and a small number of additional pipeline registers making up the remainder of the overhead.
The Zeno architecture adds a 9.8\% overhead in FPGA area (ALM Utilization) and a 5.4\% overhead in total registers relative the baseline RISC-V rv64ima system.
The synthesized configuration has an 4.3\% BRAM memory overhead.
The absolute difference in area is 4,459 ALMs, representing approximately 2\% of the total FPGA area.
The Zeno core and memory hierarchies achieve an Fmax of 130.2MHz, compared to 131.6MHz for the baseline BRISC-V core and cache hierarchy.

\begin{figure}[t]
    \centering
    \includegraphics[width=0.95\columnwidth]{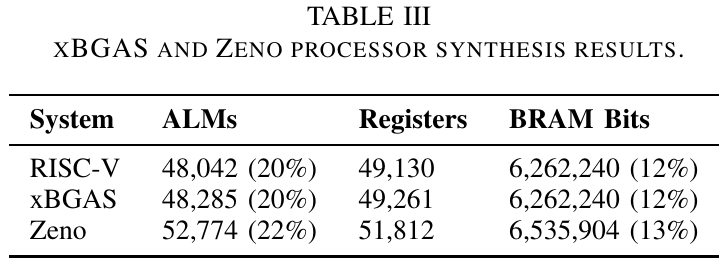}
    \label{fig:table3}
    \vspace{-0.2in}
\end{figure}

\section{Conclusion}

In this work we introduced the Zeno architecture, a scalable capability architecture for high-performance and datacenter-scale systems.
Zeno addresses the challenges of extending a capability model to a multi-node system with shared memory between nodes using a Namespace-based capability model.
Our evaluation has shown that the area overhead of the Zeno architecture is just 9.8\% that of a baseline RISC-V system.
System simulations demonstrate that application performance is not limited by capability overheads as systems are scaled up.
Our analysis of Zeno Namespace capabilities includes an evaluation of the security and performance tradeoff with an increasing number of Namespaces.
The Zeno architecture specifications, synthesized design, system support software, and programming tools will be open-sourced.

%%%%%%% -- PAPER CONTENT ENDS -- %%%%%%%%

%%%%%%%%% -- BIB STYLE AND FILE -- %%%%%%%%
\bibliographystyle{IEEEtran}
\bibliography{paper}
%%%%%%%%%%%%%%%%%%%%%%%%%%%%%%%%%%%%

\end{document}